\documentclass[aps,prl,twocolumn,superscriptaddress]{revtex4}
\usepackage{amsmath}
\usepackage[colorlinks]{hyperref}
\usepackage{amssymb}
\usepackage{color}
\usepackage{graphicx,amsmath}
\hypersetup{colorlinks,citecolor=red,linkcolor=blue,urlcolor=blue}

\begin{document}

\title{Flat bands and the physics of strongly correlated Fermi
systems}
\author{V. R. Shaginyan}
\email{vrshag@thd.pnpi.spb.ru} \affiliation{Petersburg Nuclear
Physics Institute of NRC "Kurchatov Institute", Gatchina,
188300, Russia}\affiliation{Clark Atlanta University, Atlanta,
GA 30314, USA}\author{A. Z. Msezane}\affiliation{Clark Atlanta
University, Atlanta, GA 30314, USA}\author{V. A.
Stephanovich}\affiliation{Institute of Physics, Opole
University, Oleska 48, 45-052, Opole, Poland}\author{G. S.
Japaridze}\affiliation{Clark Atlanta University, Atlanta, GA
30314, USA}\author{E. V. Kirichenko}\affiliation{Institute of
Mathematics and Informatics,Opole University, Oleska 48, 45-052,
Opole, Poland}

\begin{abstract}
Some materials can have the dispersionless parts in their
electronic spectra. These parts are usually called flat bands
and generate the corps of unusual physical properties of such
materials. These flat bands are induced by the condensation of
fermionic quasiparticles \cite{ks,phys_rep,book}, being very
similar to the Bose condensation. The difference is that
fermions to condense, the Fermi surface should change its
topology \cite{ks,vol,phys_rep,book}, leading to violation of
time-reversal (T) and particle-hole (C) symmetries. Thus, the
famous Landau theory of Fermi liquids does not work for the
systems with fermion condensate (FC) so that several
experimentally observable anomalies have not been explained so
far. Here we use FC approach to explain recent observations of
the asymmetric tunneling conductivity in heavy-fermion compounds
and graphene \cite{steg11,steg17,graph} and its restoration in
magnetic fields, as well as the violation of Leggett theorem
\cite{legget,leg2}, recently observed experimentally
\cite{bosovic,zaanen} in overdoped cuprates, and recent
observation of the challenging universal scaling connecting
linear-$T$-dependent resistivity to the superconducting
superfluid density \cite{lin}.
\end{abstract}

\maketitle

\section{Introduction}

Many strongly correlated Fermi systems like heavy-fermion (HF)
metals and high-$T_c$ superconductors (HTSC) can be viewed as
those where the familiar description in terms of the Landau
Fermi liquid (LFL) theory fails. Instead, these systems
demonstrate the so-called non-Fermi liquid (NFL) properties,
which exhibit the violation of the time-reversal (T-invariance)
and the particle-hole (antiparticle) (C) invariance. A powerful
experimental tool to detect these violations is  a tunneling
conductance. It is common knowledge that LFL theory preserves
both T and C symmetries so that the differential conductivity
$dI/dV$, $I$ being the current and  $V$ the bias voltage, is a
symmetric function of $V$. It has been predicted that the
conductivity becomes asymmetric for HF metals like $\rm
CeCoIn_5$ and $\rm YbRh_2Si_2$ whose electronic system undergoes
special transformation in a way that some part of their spectrum
becomes dispersionless, forming the so-called flat bands
\cite{tun,tun_k,pla_2007,phys_rep}. These flat bands indicate
that a system is close to a special quantum critical point,
which is a topological fermion condensation quantum phase
transition (FCQPT) \cite{phys_rep}. An important point here is
that the application of magnetic field $B$ restores the normal
Fermi-liquid properties so that the above mentioned conductivity
becomes a symmetric function of $V$ due to the reappearance of
the T- and C-invariances \cite{tun,pla_2007,phys_rep}. This
behavior has been observed in recent measurements of tunneling
conductivity on both $\rm YbRh_2Si_2$ \cite{steg11,steg17} and
graphene \cite{graph}. Our analysis of the data
\cite{steg17,graph} demonstrates that indeed the application of
magnetic field restores the symmetry of the differential
conductivity.

In many common superconductors the density $n_s$ of
superconductive electrons is equal to the total electron density
$n_t$, a manifestation of the well-known Leggett's theorem
\cite{legget,leg2}. Recent measurements on overdoped copper HTSC
oxides have demonstrated very uncommon feature $n_s\ll n_t$. In
other words, the density of paired (superconductive) charge
carriers turns out to be much lower than that predicted by the
classic Bardeen, Cooper and Schrieffer (BCS) theory, where $n_s$
is directly proportional to the critical temperature $T_c$ over
a wide doping range \cite{bosovic,zaanen}. One of the
explanations for the BCS theory violation was the electronic
system of a substance is no more normal Fermi liquid. Rather,
beyond the above FCQPT point, some charge carriers form the
so-called fermion condensate (FC), having very exotic properties
\cite{phys_rep,book}. Among them is the Leggett theorem
violation with  $n_s<<n_t$ \cite{khod97}. Indeed, in this case
the Leggett theorem does not apply since the T- and C-
invariance are violated in the systems with FC
\cite{phys_rep,book,pccp}.

There are challenging experimental observations, indicating the
universal scaling:
\begin{equation}\label{hzo}
\frac{d\rho}{dT} \propto \lambda_0^2
\end{equation}
($\rho$ is the resistivity and $\lambda_0$ is the zero temperature
London  penetration depth)  for large number of strongly correlated
HTSC's \cite{lin}. The scaling spans several orders of magnitude
in $\lambda$, signifying the robustness of the law \eqref{hzo}.
The above scaling relation is similar to that observed in Ref.
\cite{bruin} and explained in Ref. \cite{quasi}. Here we show
that the scaling is simply explained by accounting for emerging
flat bands generated by FC \cite{ks,vol}.

\section{Asymmetric conductivity and the NFL behavior}

Direct experimental studies of quantum phase transitions in HTSC
and HF metals are of great importance for understanding the
underlying physical mechanisms, responsible for their anomalous
properties.  However the experimental studies of HF metals and
HTSC are difficult since the corresponding critical points are
usually concealed in the presence of the other phase transitions
like the antiferromagnetic (AF) and/or superconducting. Recently
the challenging properties of tunneling conductivity manifesting
themselves in the presence magnetic field were observed in a
graphene with a flat band \cite{graph} as well as in HTSC and
$\rm YbRh_2Si_2$ HF metal \cite{steg11,steg17}. Studying and
analyzing these properties will shed  light on the nature of the
quantum phase transition in these substances. Most of  the
experiments on HF metals and HTSC explore their thermodynamic
properties, however, it is equally important to test the other
properties of these strongly correlated systems like
quasiparticle occupation numbers $n(p,T)$ as a function of
momentum $p$ and temperature $T$. These properties are not
linked directly to the density of states (DOS)
$N(\varepsilon=0)$ ($\varepsilon$ is a quasiparticle energy
spectrum) or to the behavior of the effective mass $M^*$. The
scanning tunneling microscopy (see e.g. Ref. \cite{harr,guy})
and point contact spectroscopy \cite{andr} being sensitive to
both the density of states and quasiparticle occupation numbers
are ideal tools to explore the effects of C and T symmetry
violation. When the  C and T symmetries are violated, the
differential tunneling conductivity and dynamic conductance are
no more symmetric function of applied voltage $V$. This
asymmetry can be observed both in normal and superconducting
phases of the above strongly correlated substances. As in normal
Fermi liquids, the particle-hole symmetry is kept intact, the
differential tunneling conductivity and dynamic conductance are
symmetric functions of $V$. Thus, the conductivity asymmetry is
not observed in conventional metals, especially at low
temperatures.

\begin{figure}[!ht]
\begin{center}
\includegraphics[width=0.5\textwidth]{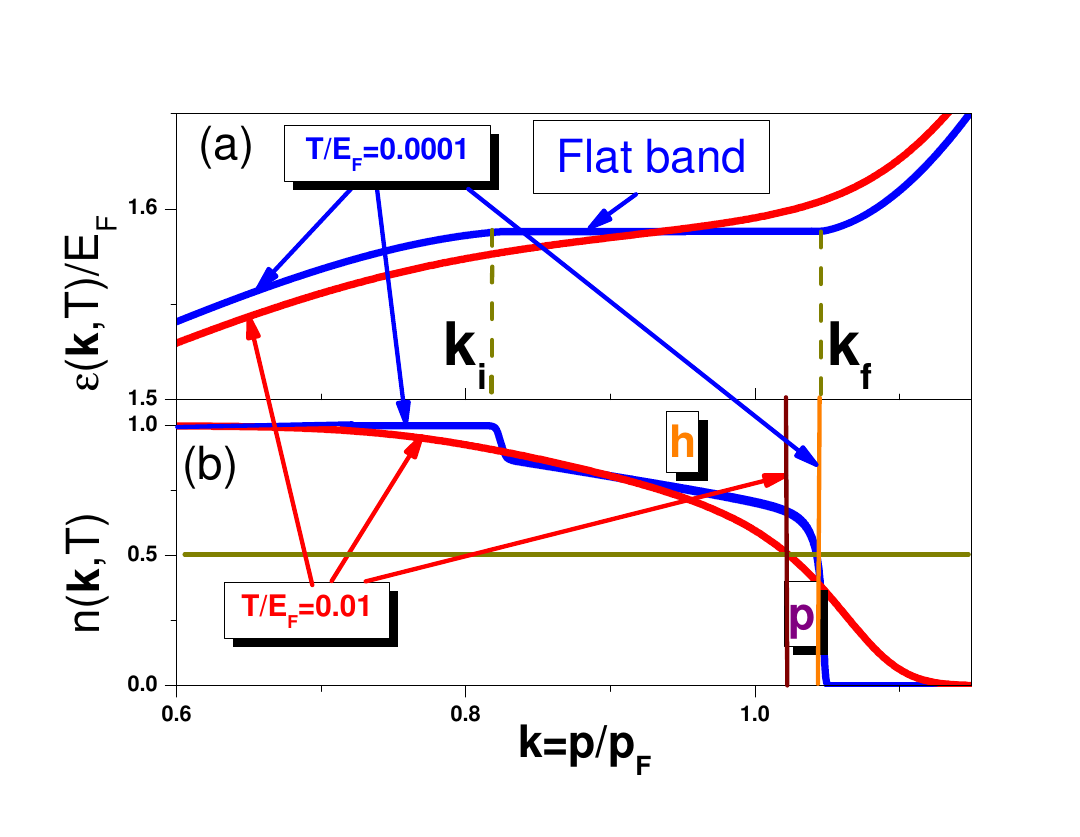}
\end{center}
\caption{Flat band induced by FC. The single particle spectrum
(a) and the quasiparticle occupation number (b) at small, but
finite temperatures versus the dimensionless momentum $k=p/p_F$,
where $p_F$ is the Fermi momentum. Temperature is measured in
the units of $E_F$. At $T=0.01E_F$ and $T=0.0001E_F$ the
vertical lines show the position of the Fermi level $E_F$, at
which $n({\bf k},T)=0.5$, see the horizontal line in panel (b).
At $T=0.0001E_F$ (blue curve), the single particle spectrum
$\varepsilon({\bf k},T)$ is almost flat (marked "Flat band") in
the range $k_f-k_i$ ($k_i$ and $k_f$ stand for initial and final
momenta for FC realization respectively, with $k=p/p_F$) so that
the distribution function $n({\bf k},T)$ becomes more asymmetric
with respect to the Fermi level $E_F$, generating the NFL
behavior and both C- and T-invariance break. To illuminate the
asymmetry, the area occupied by holes in panel (b) is marked by
letter "{\bf h}", while that for quasiparticles is marked by
letter "{\bf p}".}\label{Fig1}
\end{figure}

To determine the tunneling conductivity, we should first
calculate the tunneling current $I$ through the point contact
between two metals as a function of the voltage $V$. The best
way to calculate $I(V)$ is to use the method of Harrison
\cite{harr,guy}, in which the tunneling current is proportional
to the particle transition probability, introduced by Bardeen
\cite{bard}. Namely, Bardeen considers a probability $P_{12}$ of
a particle (say electron) transition from a state 1 on one side
of the tunneling layer to a state 2 on the other side $P_{12}
\sim |t_{12}|^2N_2(0)n_1(1-n_2)$. Here $t_{12}$ is the
transition matrix element, $N_2(0)$ is the density of states (at
$\varepsilon=0$, see above) in the state 2 and $n_{1,2}$ are
electron occupation numbers in the states 1 and 2 respectively.
Then, the total tunneling current $I$ is proportional to the
difference between those from 1 to 2 and from 2 to 1 so that
\begin{eqnarray}
I \sim P_{12}-P_{21} \sim |t_{12}|^2 N_1(0)N_2(0) \times \nonumber \\
\big[n_1(1-n_2)-n_2(1-n_1)\big]=\nonumber \\
|t_{12}|^2 N_1(0)N_2(0)(n_1-n_2).\label{hzo1}
\end{eqnarray}
The matrix element $t_{12}$ was calculated using the WKB
(Wentzel-Kramers-Brillouin) approximation \cite{harr}, yielding
$t_{12}=t(N_1(0)N_2(0))^{-1/2}$, where $t$ is the transition
amplitude. Multiplying the above expression by 2 to account for
the electronic spin and integrating over energy $\varepsilon$
leads to the following result \cite{harr,guy}:
\begin{equation}\label{ui1}
I(V)=2|t|^2\int\left[n_F(\varepsilon-V)-
n_F(\varepsilon)\right]d\varepsilon.
\end{equation}
Here $n_F(\varepsilon)$ is the electron occupation number for a
metal without FC. In the expression \eqref{ui1}, we use the
atomic units $e=m=\hbar =1$, where $e$ and $m$ are electron
charge and mass, respectively. Since temperature is low,
$n_F(\varepsilon)$ can be approximated by the step function
$\theta(\varepsilon-\mu)$, where $\mu$ is the chemical
potential. It follows from Eq. (\ref{ui1}) that quasiparticles
with the single particle energy $\varepsilon$, $\mu-V\leq
\varepsilon\leq \mu$, contribute to the current, and $I(V)=c_1V$
and $\sigma_d(V)=dI/dV=c_1$, with $c_1=const$. Thus, in the
framework of LFL theory the differential tunneling conductivity
$\sigma_d(V)$, being a constant, is a symmetric function of the
voltage $V$, i.e.  $\sigma_d(V)= \sigma_d(-V)$. In fact, the
symmetry of $\sigma_d(V)$ holds if the C and T symmetries are
observed, as it is customary for LFL theory. Therefore, the
existence of the symmetric $\sigma_d(V)$ is quite obvious and
common in case of contact of  two ordinary (i.e. those without
FC) metals, regardless are they in normal or superconducting
states.

\begin{figure} [! ht]
\begin{center}
\includegraphics [width=0.47\textwidth] {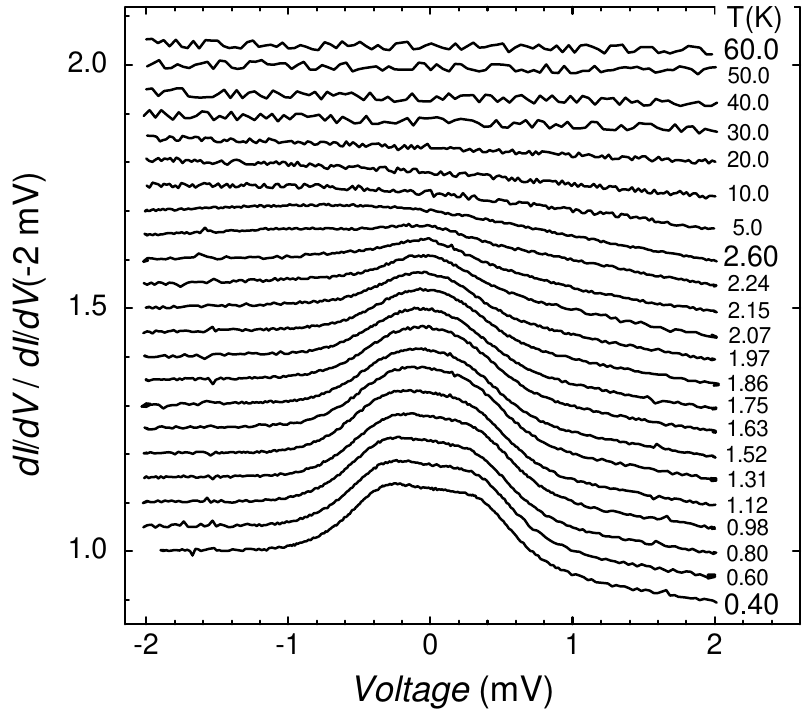}
\end {center}
\caption{Conductivity spectra $\sigma_d(V)=dI/dV$ measured on
the HF metal $\rm CeCoIn_5$ with point contacts (Au/CeCoIn$_5$)
over wide temperature range \cite{park}. Curves $\sigma_d(V)$
are shifted vertically by 0.05 for clarity and normalized by the
conductance at -2 mV. The asymmetry develops at $T\simeq 40$ K.
It becomes stronger at decreasing temperature and persists
below $T<T_c$ in the superconducting state \cite{park}.} \label
{Fig2}
\end{figure}

The situation becomes quite different if we are dealing with a
strongly correlated Fermi system locating near FCQPT, forming
flat bands \cite{ks,vol} and violating the C and T symmetries
\cite{phys_rep,book}. Figure \ref{Fig1}(a) reports the
low-temperature single-particle energy spectrum
$\varepsilon({\bf k},T)$. Figure  \ref{Fig1}(b) displays the
occupation numbers $n({\bf k},T)$ of a system with FC and shows
that the flat band generated by FCQPT is indeed accompanied by
the violation of C and T symmetries, which is reflected by the
asymmetry in the regions occupied by particles (marked {\bf p})
and holes (marked {\bf h}) \cite{phys_rep}. In case of the
strongly correlated Fermi system with FC, the tunneling current
becomes \cite{tun,tun_k,phys_rep}
\begin{equation} \label{ui8}
I(V)=2|t|^2\int\left[n(\varepsilon-V,T)-
n_F(\varepsilon,T)\right]d\varepsilon.
\end{equation}
Here one of the distribution functions of ordinary metal $n_F$
in the right side of Eq. \eqref{ui1} is replaced by
$n(\varepsilon,T)$, shown in Fig. \ref{Fig1} (b). As a result, the
asymmetric part of the differential conductivity
$\Delta\sigma_d(V)=\sigma_d(V)-\sigma_d(-V)$ becomes finite and
we obtain \cite{tun,pla_2007,tun_k,phys_rep,book}
\begin{equation} \label{ui11}
\Delta \sigma_d(V)\simeq
c\left(\frac{V}{2T}\right)\frac{p_f-p_i}{p_F}.
\end{equation}
In (\ref{ui11}) $p_F$ is the Fermi momentum, $c$ is a constant
of the order of unity. Note that the conductivity $\Delta
\sigma_d(V)$ remains asymmetric also in the superconducting
phase of either HTSC or HF metals. In that case it is again the
occupation number $n({\bf p})$ that is responsible for the
asymmetric part of $\Delta \sigma_d(V)$, since this function is
not appreciably disturbed by the superconductive pairing. This
is because the superconductive pairing is usually weaker then
Landau one, forming the function $n({\bf p})$ \cite{phys_rep}.
As a result, $\Delta \sigma_d(V)$ remains approximately the same
below superconducting $T_c$ \cite{phys_rep,shag}. It is seen
from the Eq. \eqref{ui11} and Fig. \ref{Fig2} that at raising
temperatures the asymmetry diminishes and finally vanishes at
$T\geq 40$ K. Such a behavior has been observed in measurements
on the HF metal $\rm CeCoIn_5$ \cite{park}, displayed in Fig.
\ref{Fig2}.
Under the application of magnetic field $B$ at sufficiently low
temperatures $k_BT\lesssim \mu_BB$, where $k_B$ and $\mu_B$ are
the Boltzmann constant and the Bohr magneton, the strongly
correlated Fermi system transits from NFL to the LFL regime
\cite{phys_rep,pog2002}.  As we have seen above, the asymmetry
of the tunneling conductivity vanishes in the LFL state
\cite{tun,pla_2007,tun_k,phys_rep}.

\begin{figure}[!ht]
\begin{center}
\includegraphics[width=0.5\textwidth]{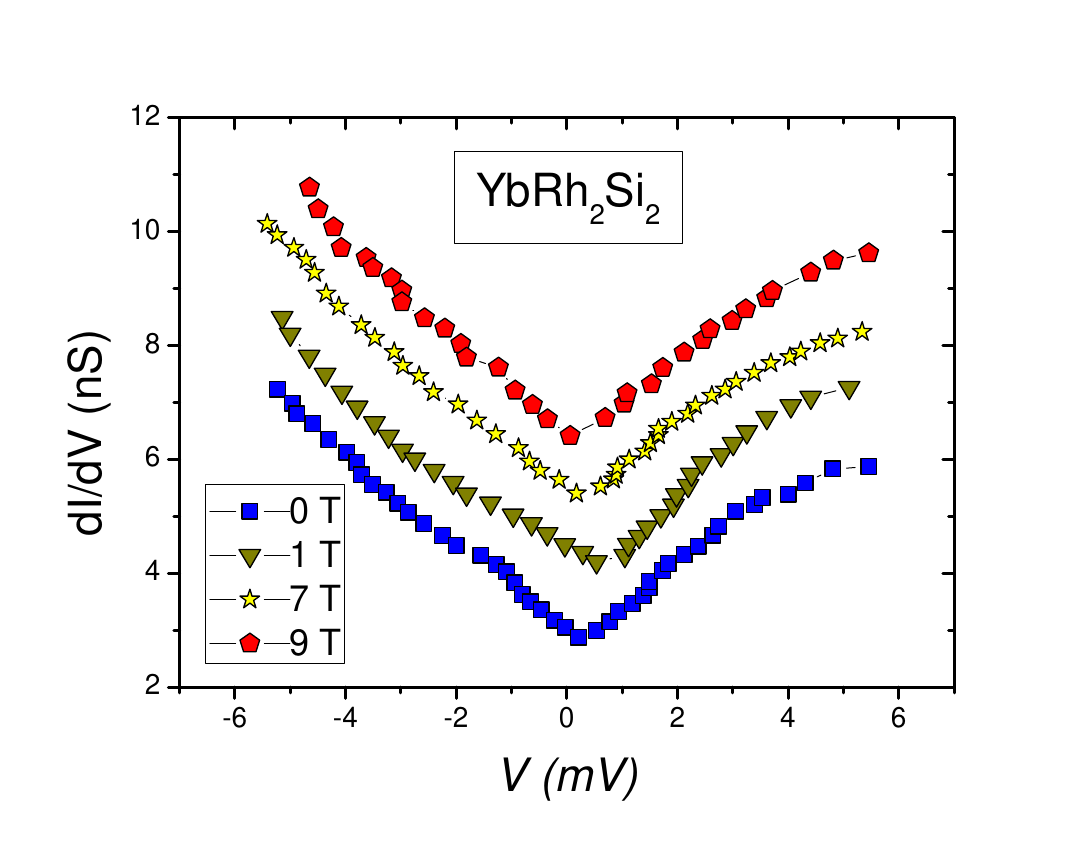}
\end{center}
\caption{Differential conductivity $\sigma_d(V)=dI/dV$ measured
on $\rm YbRh_2Si_2$ under the application of magnetic field
(legend) along the hard magnetization direction \cite{steg17}.}
\label{Fig3}
\end{figure}

\begin{figure} [! ht]
\begin{center}
\includegraphics [width=0.47\textwidth] {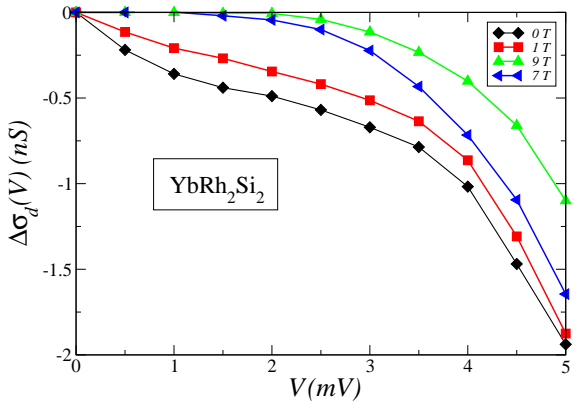}
\end {center}
\caption{The asymmetric parts $\Delta\sigma_d(V)$ of the
tunneling differential conductivity measured on $\rm YbRh_2Si_2$
and extracted from the data shown in Fig. \ref{Fig3}.} \label
{Fig4}
\end{figure}

Figure \ref{Fig3} shows the differential conductivity $\sigma_d$
observed in measurements on $\rm YbRh_2Si_2$
\cite{steg11,steg17}. It is seen that its asymmetry diminishes
at elevating magnetic field $B$, for the minima of the curves
shift to the $V=0$ point, see Fig. \ref{Fig4} for details. The
magnetic field is applied along the hard magnetization
direction, $B\parallel c$, with $B_c\simeq 0.7$ T \cite{steg17}.
Here $B_c$ is the critical field, suppressing the AF order
\cite{steg2002}. The asymmetric part of
the tunneling differential conductivity, $\Delta \sigma_d(V)$,
is displayed in Fig. \ref{Fig4}, and is extracted from
measurements shown in Fig. \ref{Fig3}. It is seen that $\Delta
\sigma_d(V)$ decreases with  $B$ increasing. We predict that the
application of magnetic field in the easy magnetization plane,
$B\bot c$ with $B_c\simeq 0.06$ T, leads to a stronger
suppression of the asymmetric part of the conductivity. This is
because in this case magnetic field effectively suppresses the
antiferromagnetic order and the NFL behavior.

\begin{figure} [! ht]
\begin{center}
\includegraphics [width=0.47\textwidth] {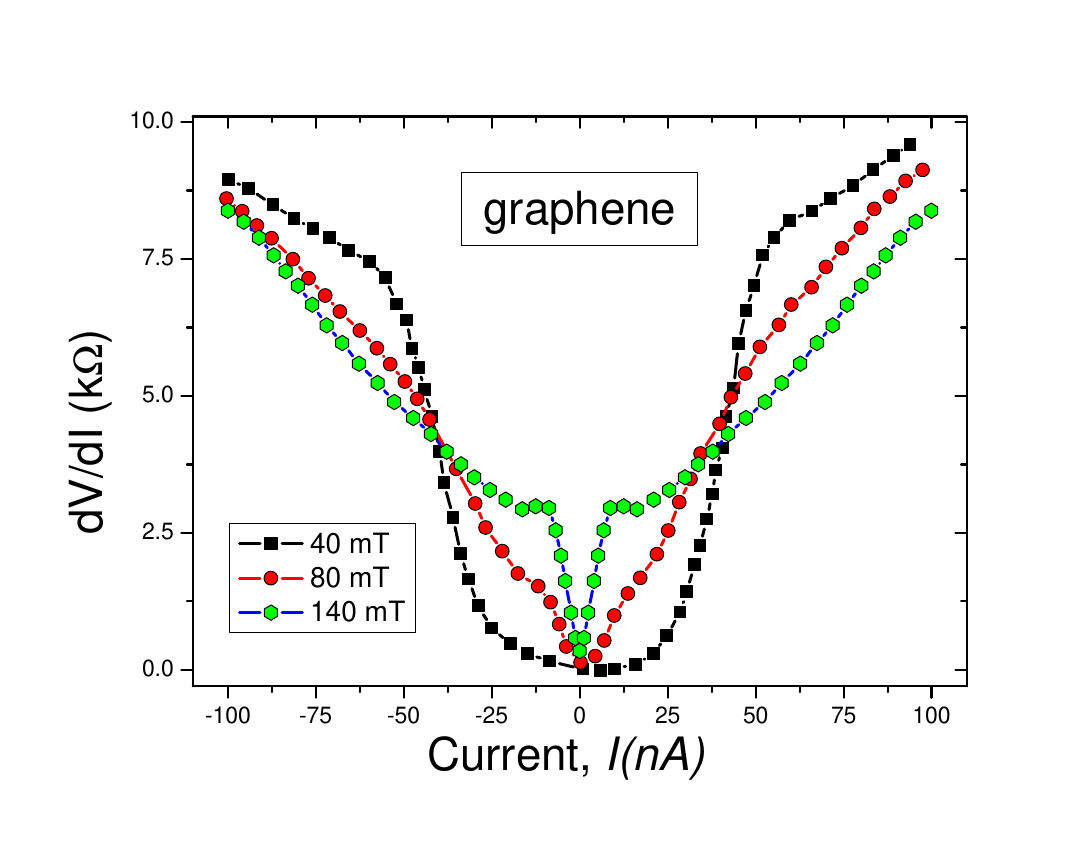}
\end {center}
\caption{Differential resistance $dV/dI$ of graphene versus
current $I$ at different magnetic fields $B$ shown in the legend
\cite{graph}. An asymmetry is seen at small magnetic fields.}
\label{Fig5}
\end{figure}

\begin{figure} [! ht]
\begin{center}
\includegraphics [width=0.47\textwidth] {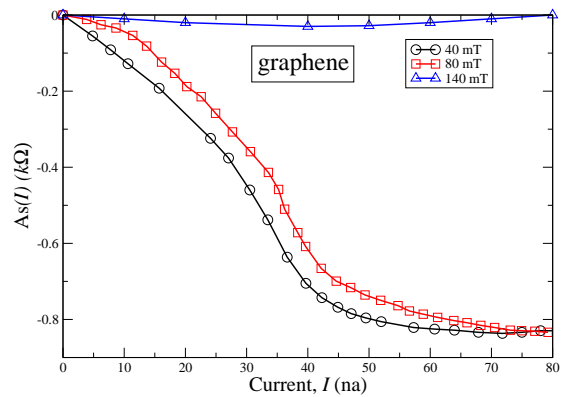}
\end {center}
\caption{Magnetic field (legend) dependence of the asymmetric
part $dV/dI(I)-dV/dI(-I)$ versus the current $I$, extracted from
the data of Fig. \ref{Fig5} for graphene.} \label{Fig6}
\end{figure}

One can measure the differential resistance $dV/dI$ as a
function of current $I$. In fact, the symmetry properties of
this function is the same as those of $\sigma_d(V)$: under the
application of a magnetic field the asymmetry of the
differential resistance vanishes as the system transits to a
normal Fermi liquid state. The differential resistance $dV/dI$
of graphene as a function of a direct current $I$ for different
magnetic fields $B$ is reported in Fig. \ref{Fig5} \cite{graph}.
Figure \ref{Fig5} shows that the asymmetric part of the
differential resistance $As(I)=dV/dI(I)-dV/dI(-I)$ diminishes at
elevated magnetic field, and vanishes at $B\simeq 140$ mT, as it
is seen from Fig. \ref{Fig6}. Such a behavior is extremely
important since the strongly correlated graphene has a perfect
flat band, meaning that the FC effects should be clearly
manifested in this material \cite{graph}.
Thus, in accordance with the prediction
\cite{tun,pla_2007,tun_k,phys_rep} the asymmetric part tends to
zero at sufficiently high magnetic fields, as it is seen from
Fig. \ref{Fig6}. The asymmetry persists in the superconducting
state of graphene \cite{graph} and is suppressed at $B\simeq
140$ mT.

Disappearance of the asymmetric part of the differential
conductivity in Fig. \ref{Fig6} shows that as the magnetic field
increases, graphene transits from the NFL  to the LFL state. To
support this statement, we surmise that the resistance $\rho(T)$
should exhibit the linear dependence $\rho(T)\propto T$ in the
normal state at zero magnetic field, while at elevated magnetic
fields and low temperatures $k_BT<<\mu_BB$, the resistance
becomes a quadratic function of temperature $\rho(T)\propto T^2$
as it is in the case of the strongly correlated Fermi systems
\cite{phys_rep,book,khod2012}.

\section{Overdoped cuprates}

Recent striking experimental results of Bo\^zovi\'c {\em et al}
\cite{bosovic} can be regarded as one  more manifestation of
violation of the time-reversal (T) invariance and the Leggett
theorem \cite{legget} in the systems with the flat bands,
associated with the FC state. To be specific, the measurements
revealed that the superconducting phase transition temperature
$T_c$ in the overdoped copper oxides La$_{2-x}$Sr$_x$CuO$_4$
(LSCO) is linearly proportional to the 2D superfluid phase
stiffness $\rho_s(T=0)=\rho_{s0}$ \cite{bosovic}. This unusual
dependence cannot be described by the standard BCS theory as,
for instance, the density of paired charge carriers, forming the
superfluid density $n_s$ and proportional to density $\rho_{s0}$
does not equal to the total particle (charge carriers in our
case) density $n_{el}$. This contradicts the Leggett theorem
\cite{legget}, stating  that $n_s \sim n_{el}$ for any
superfluid liquid at $T=0$. The Leggett theorem is valid as long
as system is C and T invariant, see, e.g. Refs.
\cite{legget,leg2}. Since FC state, being highly topologically
nontrivial \cite{book,tun,phys_rep}, violates both C and T
symmetries, the equality $n_s\simeq n_{el}$ is no longer valid
for it. The approximate equality above is replaced by the
inequality $n_s=n_{FC}\ll n_{el}$, where $n_{FC}$ is the density
of particles in FC state \cite{pccp}.  This implies that the
main contribution to the unusual properties of the above
superconductive overdoped copper oxides comes from the FC state.
We conclude that the underlying physical mechanism of both the
unusual properties of overdoped copper oxides and above
asymmetry in tunneling conductivity is a fermion condensation
phenomenon, see e.g. \cite{phys_rep,ks,vol,book}.

\section{Universal scaling relation}

Another experimental result \cite{lin} providing insight deals
with the universal scaling relation, which can also be explained
using the flat band concept. The authors \cite{lin} measured
$d\rho/dT$ ($\rho$ is a resistivity) for a large number of HTSC
substances (with LSCO and well-known HF compound CeCoIn$_5$
among them, see Table I of Ref. \cite{lin}) for $T>T_c$ and
discovered the remarkable result. Namely, $d\rho/dT$ for all
mentioned substances depends linearly on $\lambda^2 (T=0)\equiv
\lambda_0^2$, where $\lambda$ is the London penetration depth.
Note that all  the above superconductors belong to the London
type since for them $\lambda
>> \xi_0$, where $\xi_0$ is the zero temperature
coherence length, see, e.g. Ref. \cite{pccp}.

It  has been shown in Ref. \cite{lin}, that the scaling relation
\begin{equation}\label{cs1}
\frac{d\rho}{dT} \propto \frac{k_B}{\hbar}\lambda_0^2
\end{equation}
spans over several orders of magnitude in $\lambda_0$ signifying
the robustness of the above law. At the phase transition point
$T=T_c$, relation \eqref{cs1} yields the well-known Holmes law
(Ref. \cite{lin}, see also Ref. \cite{kogan} for its theoretical
derivation)
\begin{equation}\label{cs2}
\sigma T_c \propto \lambda_0^{-2},
\end{equation}
where $\sigma = \rho^{-1}$ is the normal state {\em dc}
conductivity. It has been shown by Kogan \cite{kogan} that even
for oversimplified model of isotropic BCS superconductor, the
Holmes law holds. In the same model of a simple metal, one can
express resistivity $\rho$ through microscopic substance
parameters \cite{agd}: $e^2n\rho \simeq p_F/(\tau v_F)$, where
$\tau$ is the quasiparticles lifetime, $n$ is the carrier
density, and $v_F$ is the Fermi velocity. Taking into account
that $p_F/v_F=M^*$, we arrive at the expression
\[
\rho=\frac{M^*}{ne^2\tau},
\]
which formally agrees with the Drude formula. It has been shown
in Ref. \cite{pccp} that the agreement with experimental results
\cite{bosovic} is achieved if we assume that the effective mass
and the density are attributed to the carriers in the FC state
only, i.e. $M^* \equiv M_{FC}$ and $n\equiv n_{FC}$. Keeping
this in mind and utilizing the relation $1/\tau=k_BT/\hbar$ (see
\cite{khod2013} and Chapter 9 of Ref. \cite{book}), we obtain
\begin{equation}\label{cs3}
\rho=\frac{M_{FC}}{e^2n_{FC}}\frac{k_BT}{\hbar}\equiv 4\pi
\lambda_0^2 \frac{k_BT}{\hbar},
\end{equation}
i.e. $d\rho/dT$ is indeed given by the expression \eqref{cs1}.
Equation \eqref{cs3} demonstrates the concept that fermion
condensation can explain all the above experimentally observed
universal scaling relations. Note that the presented FC approach
is not affected by the microscopic non universal features of the
substances under study. This is due to the fact that the FC
state is protected by its topological structure, representing a
new class of Fermi liquids \cite{vol,book}. Namely, the
consideration of the specific crystalline structure of a
compound, its anisotropy, defect structures etc do not change
our results qualitatively. This strongly suggests that the FC
approach is certainly viable theoretical tool to explain the
universal scaling relations similar to those discovered in the
experiments \cite{bosovic, lin}. In other words, the fermion
condensation of charge carriers in the considered strongly
correlated HTSC's, begotten by quantum phase transition,  is
indeed the primary physical mechanism for their observable
universal scaling properties. This mechanism can be extended to
a broad set of substances with a very different microscopic
characteristics \cite{phys_rep,book}.

\section{Conclusions}
The main message of the present paper is that if the electronic
spectrum of a substance by some reasons has the dispersionless
part, called flat bands, this is exactly what is responsible for
a many puzzling and nontrivial experimentally observable
features of this substance. This is notwithstanding the
different microscopic properties (like crystalline symmetry,
defect structure etc) of these substances. Explanation for that
stems from the fact that the fermion condensation most readily
occurs in the substance with the flat bands. The experimental
manifestations of FC phenomenon are varied, which means that
different experimental techniques are suitable to detect them.
We have shown that one of such techniques is the scanning
tunneling microscopy. The reason is that it is sensitive to both
the density of states and quasiparticle occupation numbers,
which makes it suitable for studying the effects related to the
violation of the particle-hole symmetry and to the time-reversal
invariance. The violation of the particle-hole symmetry and of
the time-reversal invariance leads to the asymmetry of the
differential tunneling conductivity and resistance to the
applied voltage $V$ or current $I$. Based on the recent
experimental results, we have demonstrated that the asymmetric
part of both the conductivity and the resistivity vanishes under
the application of magnetic field, as predicted in Refs.
\cite{tun,pla_2007,tun_k}. To support our statement on the
violation of the T-invariance, we analyzed and discussed within
the fermion condensation framework the recent challenging
measurements on overdoped cuprates by Bo\^zovi\'c and coauthors
\cite{bosovic} based on the fermion condensation framework. We
also described and explained, using the FC concept, the yet
unexplained observations of  scaling \cite{lin}. Finally, our
study of the recent experimental results  suggests that FCQPT is
the intrinsic feature of many strongly correlated Fermi systems
and can be viewed as the universal cause of the NFL behavior.

\section{Acknowledgement} This work was partly supported by U.S.
DOE, Division of Chemical Sciences, Office of Basic Energy
Sciences, Office of Energy Research.

\end{document}